\documentclass{mn2e}
\usepackage{graphicx}

\title [Elliptical Outflows in Quasars]{Discovery of Universal Elliptical Outflow Structures
in Radio-Quiet Quasars}
\author[J. Lovegrove, R. Schild \& D. Leiter]{Justin Lovegrove$^{1,2}$\thanks{Corresponding author, email jl805@soton.ac.uk}, Rudolph E. Schild$^1$ and Darryl Leiter$^3$\\
$^1$Harvard-Smithsonian Center for Astrophysics, 60 Garden Street, Cambridge, MA 02138, USA\\
$^2$Schools of Mathematics and Physics \& Astronomy, University of Southampton, University Road, Southampton, SO17 1BJ, UK\\
$^3$Visiting Scientist, National Radio Astronomy Observatory, 1180 Boxwood Estate Road, Charlottesville, VA 22903-4608}
\date{Submitted 29 March 2010}

\pagerange{\pageref{firstpage}--\pageref{lastpage}} \pubyear{2010}

\begin{document}

\label{firstpage}

\maketitle

\begin{abstract}
Fifty-nine quasars in the background of the Magellanic Clouds had
brightness records monitored by the MACHO project during the years
1992 - 99. Because the circumpolar fields of these quasars had no
seasonal sampling defects, their observation produced data sets well
suited to further careful analysis. Following a preliminary report
wherein we showed the existence of reverberation in the data for one
of the radio-quiet quasars in this group, we now show that similar
reverberations have been seen in all of the 55 radio-quiet quasars with
adequate data, making possible the determination of the quasar inclination
to the observer's  line of sight. The reverberation signatures indicate the
presence of large-scale elliptical outflow structures
similar to that predicted by the Elvis
(2000) and ``dusty torus'' models of quasars, whose characteristic sizes
vary within a surprisingly narrow range of scales. More importantly
the observed opening angle relative to the polar axis of the
universal elliptical outflow structure present was
consistently found to be on the order of 78 degrees.

\end{abstract}
\section{Introduction}

The wide-field imaging MACHO project produced brightness curves for
approximately 6.5 million stars (Alcock et al, 1999), among which are
brightness curves of 59 background quasars. An initial analysis of these
(Schild, Lovegrove and Protopapas, 2009; PaperI) showed that these are well
suited for systematic analysis because the Magellanic Clouds are circumpolar
and the data do not suffer seasonal dropouts from when the quasar was too
close to the sun.

The initial analysis was inspired by the fact that the 25-year brightness
record of the gravitationally lensed Q0957 quasar had shown evidence of
reverberation structure (Thomson and Schild, 1997), that was easily
detected in autocorrelation. The brightness amplitudes were typically 0.2
mag and the reverberations were on time scales of several hundred days.
Because the pulse trains were frequently overlapping, it was difficult to
recognize the pulse train from simple inspection.

In our Paper I initial analysis we chose a single quasar, MACHO 13.5962.237,
and demonstrated that the structure found in autocorrelation was
realized by averaging together multiple wave trains, and the wave trains
showed multiple peaks as had been already inferred from Q0957, and
interpreted successfully by Schild (2005) as reverberations from the multiple
surfaces that occur in the region of the dusty torus. Such structure had
already been inferred by Elvis (2000) from a comprehensive analysis of a great
wealth of published spectroscopic quasar data.

A surprising discovery was that in addition to positive reverberations, in
which an initial narrow brightening pulse is followed after approximately
80 days by several broader
pulses, an almost equal number of $fading$ initial pulses with comparable
widths and lags were found. A key finding was that the UV-optical
reverberations originate at the same radial distance as the dusty torus
already inferred to exist from emission line and infrared continuum
reverberation.

The existence of luminous outer structure was also inferred from
microlensing studies of quasars, where failure of the standard luminous
accretion disc model of Shakura and Sunyanaev (1973; S-S)
to reproduce the observed
microlensing in Q2237 has been a persistent problem.

The first attempt to simulate the Q2237 microlensing with a S-S disc by
Wyithe, Webster and Turner (2001) produced a prediction for a large
microlensing brightening event that was to occur in 2001 but
never happened.
In a next attempt, Kochanek (2004) applied the model, but found that it
could work only by seizing upon highly improbable intrinsic quasar
brightening in coincidence with the microlensing. It was also noted that
for expected cosmological velocities the inferred masses of the microlenses
were sub-stellar, and like the Wyithe simulation, it predicted (Fig 10) the
existence of 2.5 magnitude microlensing events, which have never been
observed.
A final attempt to apply the model by Eigenbrod et al (2008) encountered
the same problems, and introduced a bias factor as a prior to force
the calculation to fit the improving data set with stellar mass
microlenses, but again predicted 2.5
magnitude microlensing events that have never been observed.

A successful simulation by Vakulik et al (2007) abandoned the S-S disc model
and simply allowed the calculation to find the size parameters describing
the diameter of the microlensed inner
structure, the masses of the microlenses,
and the fraction of the total luminosity responsible for the observed
brightness curves. These are the physical parameters to which the Q2237
microlensing would be most sensitive.

This objective simulation produced a successful fit to the data with
no bias factors, and with events
that were highly probable from a quasar having only 1/3 of its UV-optical
continuum radiation originating in the central region, within
6 gravitational radii, and
the remaining 2/3 in a larger outer structure. The model predicts smaller
amplitude microlensing events as observed, and determines that for expected
cosmological velocities the microlenses have planetary mass. It had
previously been demonstrated by Schild and Vakulik (2003) that such
a population of microlenses was necessary to explain the rapid
microlensing seen in another lensed quasar, Q0957.

Additional microlensing results have hinted at extended UV-optical
emission. Analysis of quadruple-image lenses by Pooley et al
(2007) showed that the optical emission comes from a
region approximately $ 3 \times 10^{16}$ cm in radius,
although this is ambiguous if
the emitting region is an extended ring with inner radius and thickness
dimensions, especially since this size was also inferred by SLR06 from
estimates of the ring thickness (Fig. 1; $ \delta r = 2 \times 10^{16} cm$).
Microlensing of the emission line region of SDSS 1004+4112 by Richards et
al (2004) demonstrated that the characteristic size of the
emitting region must be of order $10^{16} cm$, based upon the duration of
an emission line
microlensing event. We propose that the dusty torus, the broad-emission
line region, the UV-optical continuum reverberation region, and the
infrared reverberation region are all the same outer quasar structure.
Henceforth we simply refer to it as the dusty torus.
The central radius to this
structure is of order $10^{17.3} cm$, its thickness parameter
is $10^{16.3} cm$, and the height of of the luminous ring above and below
the accretion disc plane
is $10^{16.7} cm$. A cross-sectional
view of the quasar with this picture is given as Fig. 1 of SLR06.
The microlensing signature of such a structure has been simulated by Schild
\& Vakulik (2003) and by Abajas et al (2007).

The preponderance of microlensing and reverberation evidence now supporting
the dusty torus model of quasar structure now challenges the
theoretical community. The Dusty Torus model was comprehensively discussed in
1993, (Antonucci 1993) but as yet has no explanation in a standard model
of quasar
structure. However a model proposed to explain the existence of the
different observed quasar spectral states (radio loud - radio quiet)
offers a possibility. While the standard black hole structure model has a
magnetic field originating in the accretion disc, the relationship of the
accretion disk to the size and location of the dusty torus is not clear. On
the other hand Robertson and Leiter (2006; SLR06) have shown that a strong magnetic
field anchored to a compact, highly red shifted, rotating central object called
a magnetospheric eternally collapsing object (MECO) is also a viable solution to
the Einstein-Maxwell field equations.  Such an object would exhibit the effects
of a co-rotation radius in the accretion disk to explain the observed quasar
spectral states (Schild, Leiter, and Robertson, 2008; SLR08) and the location of
the dusty torus would be associated with magnetic reconnection effects generated
by the twisting of the dipole field lines to into toroidal near the light
cylinder. In addition such a model would be radiatively inefficient near the
central object’s surface as a relativistic effect of the large intrinsic
redshift there.

Because our discovery that reverberation of the UV-optical continuum
evidences important outer structure seriously challenges the standard
black hole model, our report will focus on demonstrating that $all$ $quasars$
show evidence for such outer structure. We adopt the standard definition
of quasars as luminous cosmological objects with a stellar appearance
at 1 arcsec resolution, and broad blue-shifted emission lines. Previous
work (SLR 06, 08) has made the connection that the reverberation radius is
approximately the same as the central distance of the outer region
originating the broad blue-shifted lines as an outflow wind. Our conclusion
that the Elvis flow occurs where all quasars have reverberating
luminous outer structure
would seem to prove that such quasar structure is universal.

\section {Preliminary Processing of MACHO Quasar Brightness Data}

Our preliminary processing has followed the procedures of
Schild, Lovegrove, and Protopapas (2009; Paper 1).
Raw data from standard V and R
filters was corrected for CCD defects by simple removal of any 5 sigma
data points. The data records were rebinned into uniformly spaced super-bins,
with the number of super-bins equal to the number of observations,
and all data within such a super-bin averaged. Super-bins containing no
data were linearly interpolated over. Since the original brightness records
had typically 600 data points spread over 2600 nights, a typical super-bin
has time resolution of 4 observers' days. The timing of these observations
was then rescaled for cosmological redshift. For the highest-redshift object in
the survey, MACHO 208.15799.1085 at z = 2.77 there is insufficient
observing time to produce more than one full reverberation pattern
in quasar proper time. Therefore, we
have excluded the source from further processing or
plotting. For this reason our original sample of 59 quasars less 3
radio loud objects reduces to a sample of 55 radio quiet objects.

In Paper I a quasi-periodicity with amplitude of 30 \% was found in some
but not all quasars. This variability has been called ``red-noise'' in some
contexts, but it is well observed and real and in need of explanation, but
we defer its further discussion. We remove this signature by forming
a 300-day running boxcar smoothing algorithm over the binned and
interpolated brightness record. An example of this procedure for MACHO
quasar 13.5962.237 is given as Fig. 1 of Paper I.

Following this preprocessing we have computed the autocorrelation function
separately for the V and R filter data. The autocorrelation function
always shows important structure, with a strong central peak having a
brightness amplitude of order $ 30\%$   and lags up to approximately 50 days,
followed by a
100-day string of lags with negative autocorrelation, and then several
positive autocorrelation peaks with only 10 \% autocorrelation amplitude.
In Paper 1 we have shown from a noise simulation that the peaks are real,
since a realisitc noise simulation shows noise that occurs on time scales
of our super-pixel resolution, approximately 4 observer's days, but which
we do not see in our real data. We also show in
Paper I that the autocorrelation peaks are confirmed to be real brightness
enhancements with a reverberation pattern because we can co-add data
segments containing the pattern and directly find a repeating wave form
in brightness that confirms the autocorrelation pattern.

Throughout this series of papers we presume that any structure found in
brightness records for lag $t$ reflects quasar structure on size scale
$ct$. In other words, we presume that the brightness features are caused by
structure excited in the central region and propagating outward at the
speed of light.

\section {The Mean Autocorrelation Due To Central Structure}

In Fig. 1 we show a plot of the lag of the first minimum of the
autocorrelation function separately for the V-filter and R-filter data.
As a function of this we
plot the (anti-)correlation amplitude. Based upon the theoretical error
bars associated with the autocorrelation function, we see that the formal
errors are comparable to the symbol size.

We find in Fig. 1 that all the points lie in the lag range 30 - 260 days.
The anti-correlation amplitude is between 0 and -0.5. The points scatter
around this delimited area, with no trend evident. For a typical
quasar with an anticorrelation peak lag of 100 days and an anticorrelation peak
of -.25, we infer that significant anticorrelation exists and that an upper
limit to the size of the excited central structure is 100 light days, or
$2 \times 10^{17} cm$.

\begin{figure*}
\begin{center}
\includegraphics{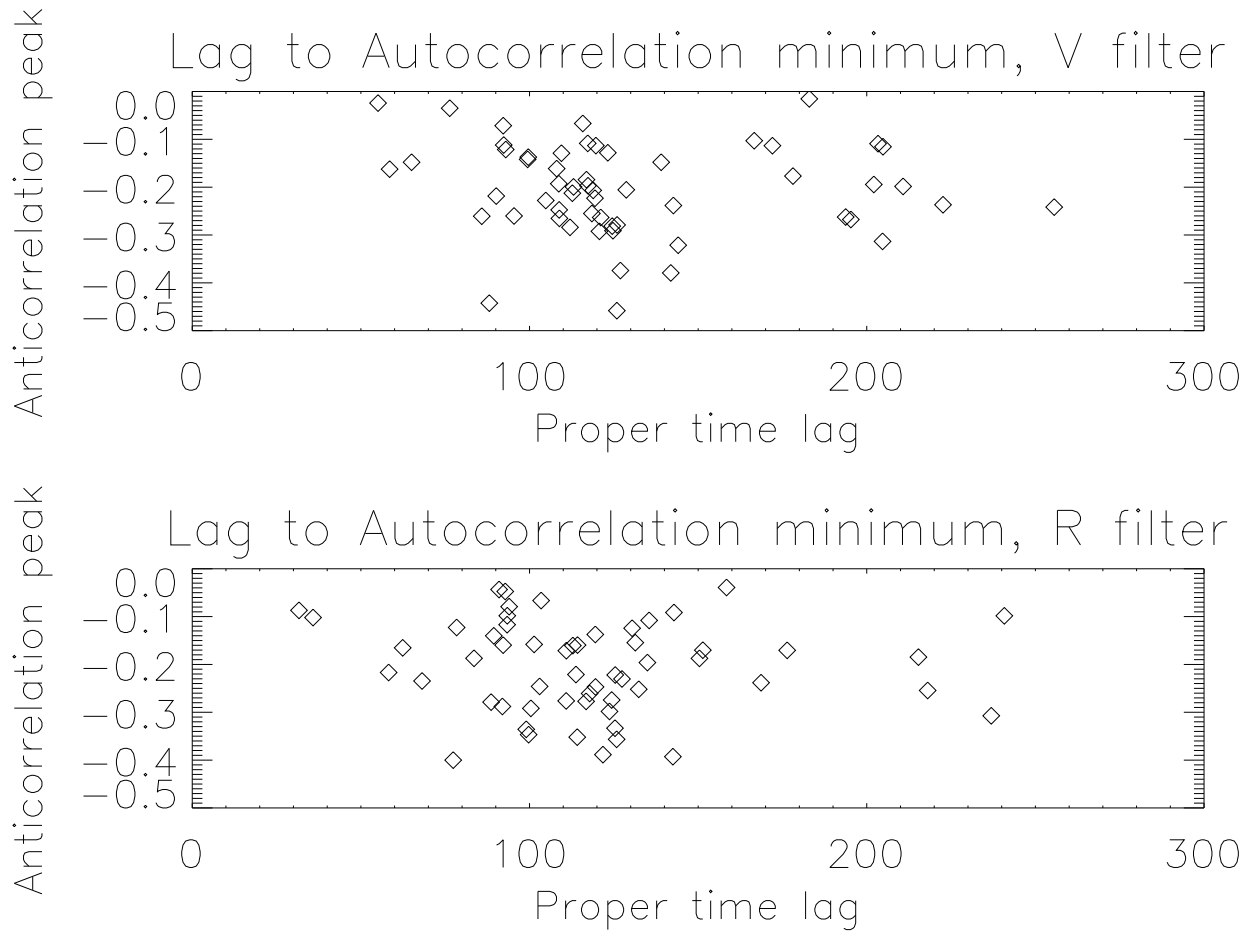}
\caption{ Autocorrelation minimum values of 55 MACHO
  quasars as a function of
  the lag time of the autocorrelation minimum (the anti-correlation maximum) for
  the V data (top) and for the R data (bottom). The anti-correlation peak
  lag seems unrelated to the amplitude of the peak.
  The anti-correlation lag is surprisingly limited to only a small range,
  less than a factor ten, particularly in the bottom plot.
  One might interpret this anti-correlation peak as an
  indicator of a dark region around the central luminous UV-optical source,
  implying that all the MACHO quasars are approximately the same size.}
\label{fig. 1}
\end{center}
\end{figure*}

In this context the siginificant result of our research was the finding
that the radio quiet MACHO quasars are quite homogenous in their observed
physical properties. It appears that the inner structures of these radio
quiet are very similar.

\section{Estimation of Quasar Viewing Angles}

Our paper I analysis of a single quasar, MACHO 13.5962.237, showed that the
pulse trains originating at the Elvis surfaces could be recognized in the
autocorrelation function and also in the wave train averaged from many data
segments averaged together in a rebinning that fixes the starting point of
each pulse sequence. This starting pulse is in principle easy to recognize,
but in practice difficult to identify because the pulse sequences overlap.
Therefore for Paper I we created the mean brightness profile from the ten most
well-defined peaks (well-defined meaning having both relatively large amplitude
and narrow width and also being resolved from the neighbouring peaks) in the
brightness record. This mean profile was then inspected and had reverberation
structure identified.

We undertook this process of hand fitting for an additional 29 quasars,
which were selected as the ones with the most complete data records.
For all of the 29 carefully studied quasars, we found that the pulse trains
seen were the same as the structure indicated by a simple
autocorrelation. However we were uncomfortable about the hand fitting and
piecemeal selection of the data, and therefore we devised a comprehensive
machine processing procedure to automate it by analyzing only the
autocorrelation function. All of the results reported herein were obtained
from the automatic machine processing.

The expected quasar reverberation pattern illustrated as Fig. 2 of Paper 1
shows that there will be 3 complications. The first is that the pattern of
reverberation at the quasar depends in a complicated way on the
orientation, such that the second reverberation peak measured would be from
the same hemisphere of the quasar for quasars seen nearly pole-on, but from
the opposite hemisphere for quasars seen nearly equator-on. This is why the
equations for the reverberation pattern originally presented in Schild (2005)
included a ``case 1'' and ``case 2''. In our present treatment, we simply
compute the lags for the four surfaces independent of viewing angle, and so
the role of the two cases reverses for inclinations above $90 - \epsilon$
degrees, where $\epsilon$ is the internal structure variable defined in Schild
(2005) as the angle between the accretion disc plane and the brightest parts
of the luminous regions of the Elvis surfaces.

A second complication arises from consideration of SLP09 Fig.2 lower panel,
where the pulse trains expected are shown. It is obvious from simple
inspection that a train with four pulses will not always be observed,
because at the extreme angles of 0 and 90 degrees and also at $90 - 
\epsilon$ degrees,
several of the pulses merge pairwise. Therefore we should expect some
quasars to be found with either 2, 3, or 4 reverberation pulses after
the initial pulse. We will call this the pulse merging problem.

Finally, we see in the figure that for small quasar viewing angles,
the measured
value of the viewing angle will be poorly determined because it is a weak,
slowly varying function of the pulse lags, and because of the pulse merging
problem.

Therefore the process of machine-computing of the inclination angle was
undertaken as follows. First we identified the 27 quasars for which the
full pulse train of four pulses was found, and computed inclination angles
for them. We noticed that the radius parameter can be easily
calculated by adding the 4 lags together and dividing by four. From the
equations for the lags given in SLP09 it can easily be seen that $t_1 + t_2
+ t_3 + t_4  = 4 R_{blr}/c$, the radius to the broad line region, and hence the
fundamental size scale of the quasar. This allows us to rescale all of the
reverberations as fractions of $R_{blr}$, so that these fractional lags would
be the same for all the quasars having the same inclination. Then from the
equations for $t_1$-$t_4$  with $R_{blr}$ known, we can compute values
for inclination angle $\theta$ and internal structure variable $\epsilon$.
The average of the two estimates for each angle and
their rms deviation give a mean value of the angle and an error statistic,
which we consistently quote as 1-$\sigma$ errors.

For the quasars with only 3 measured reverberation peaks, we initially assumed
that the pulse merging problem causes the missing lag, and determined the
inclination angle the same way except we assume that $t_2$ and $t_3$ have
merged, thus forcing $\theta = 90^o - \epsilon$.

Finally, the quasars with only 2 peaks were presumed to be double-merger
products, and because of the complication of attempting to locate the
centroid of the two peaks which will probably be asymmetrical, we have
simply entered the inclination angle as 90 degrees in our table.

This procedure was followed for all the quasars, and for the V and R
data separately. To give some sense of how well our inclination angle
measurement worked, we list both estimates in our Table 1. In this table we
list the MACHO ID, redshift z, number of V and R observations (n(V) and n(R)),
mean V and R apparent magnitudes, our estimated $R_{blr}$, estimated viewing
angle $\theta$, and the internal structure variable $\epsilon$.

\begin{table*}
\caption{Properties of the MACHO quasars.  z = redshift.  n(V) and n(R) = total number
of V and R observations respectively.  V and R = mean V and R magnitudes respectively.  $R_{blr}$
(in light days), $\theta$ and $\epsilon$ (both in degrees) are calculated parameters
based on reverberation from elliptical outflows.}
\begin{tabular}{lcccccccr}
\hline
MACHO ID & z & n(V) & V & n(R) & R & $R_{blr}$ & $\theta$ & $\epsilon$\\
\hline
2.5873.82 &  0.46  &    959 &   17.44  &  1028 &  17.00  &  608 &   71.0  &  9.5 \\
5.4892.1971 &  1.58  &    958 &   18.46  &  938 &   18.12  &  560 &   70.0  &  12.0 \\
6.6572.268 &  1.81  &    988  &  18.33  &  1011 &  18.08  &  578 &   71.0  &  12.5 \\
9.4641.568 &  1.18  &    973  &  19.20  &  950 &   18.90 &   697 &   72.0 &   11.0 \\
9.4882.332 &  0.32  &    995  &  18.85  &  966 &   18.51  &  589 &   64.5  &  12.0 \\
9.5239.505 &  1.30  &    968  &  19.19  &  1007 &   18.81  &  579 &   64.5  &  12.0 \\
9.5484.258 &  2.32  &    990  &  18.61  &  396  &  18.30   & 481 &   83.5  &  10.5 \\
11.8988.1350 &  0.33  &    969 &   19.55  &  978 &   19.23  &  541 &   67.0 &   14.0 \\
13.5717.178 &  1.66  &    915  &  18.57  &  509 &   18.20  &  572  &  59.5 &   13.0 \\
13.6805.324 &  1.72  &    952  &  19.02  &  931  &  18.70  &  594  &  85.0 &   9.5 \\
13.6808.521 &  1.64  &    928  &  19.04  &  397  &  18.74 &   510 &   81.5 &   11.0 \\
17.2227.488 &  0.28  &    445  &  18.89  &  439  &  18.58  &  608 &   82.5 &   8.0 \\
17.3197.1182 &  0.90  &    431 &   18.91 &   187 &   18.59  &  567 &   73.0 &   16.0 \\
20.4678.600 &  2.22  &    348  &  20.11 &   356  &  19.87  &  439  &  68.0 &   14.0 \\
22.4990.462 &  1.56  &    542  &  19.94 &   519  &  19.50  &  556  &  64.5 &   12.5 \\
22.5595.1333 &  1.15  &    568  &  18.60 &   239 &   18.30  &  565 &   73.5 &   9.0 \\
25.3469.117 &  0.38  &    373  &  18.09 &   363  &  17.82  &  558  &  65.0 &   12.0 \\
25.3712.72 &  2.17  &    369  &  18.62  &  365  &  18.30  &  517  &  72.0   & 12.0 \\
30.11301.499 &  0.46  &    297 &   19.46 &   279 &   19.08 &   546 &   68.5  &  12.5 \\
37.5584.159 &  0.50  &    264  &  19.48  &  258 &   18.81  &  562  &  70.5  &  12.5 \\
48.2620.2719 &  0.26  &    363  &  19.06 &   352 &   18.73 &   605  &  65.0 &   13.0 \\
52.4565.356 &  2.29  &    257  &  19.17  &  255  &  18.96  &  447  &  85.0  &  8.0 \\
53.3360.344 &  1.86  &    260  &  19.30  &  251  &  19.05 &    496 &   67.5 &   10.0 \\
53.3970.140 &  2.04  &    272 &   18.51  &  105  &  18.24  &  404  &  69.0 &   13.5 \\
58.5903.69 &  2.24  &    249 &   18.24   & 322  &  17.97  &  491  &  80.5   & 8.5 \\
58.6272.729 &  1.53  &    327 &   20.01  &  129  &  19.61  &  518 &   70.5  &  12.5 \\
59.6398.185 &  1.64  &    279  &  19.37  &  291  &  19.01  &  539 &   83.0  &  9.5 \\
61.8072.358 &  1.65  &    383  &  19.33  &  219  &  19.05  &  471 &   67.5  &  16.0 \\
61.8199.302 &  1.79  &    389  &  18.94  &  361  &  18.68  &  475 &   69.0  &  14.0 \\
63.6643.393 &  0.47  &    243  &  19.71  &  243  &  19.29  &  536 &   67.0  &  11.0 \\
63.7365.151 &  0.65  &    250  &  18.74  &  243  &  18.40  &  625 &   83.5  &  10.5 \\
64.8088.215 &  1.95  &    255  &  18.98  &  240  &  18.73  &  464 &   67.5  &  17.5 \\
64.8092.454 &  2.03  &    242  &  20.14  &  238  &  19.94  &  485 &   73.5 &   10.5 \\
77.7551.3853 &  0.85  &    1328 &  19.84  &  1421 &  19.61  &  489 &   69.0  &   14.0 \\
78.5855.788 &  0.63  &    1457 &  18.64  &  723  &  18.42   & 491 &   77.0   & 12.0 \\
206.16653.987 &   1.05  &    741 &   19.56 &   581 &   19.28  &  486 &   69.0 &   14.5 \\
206.17052.388  &  2.15  &    803 &   18.91 &   781  &  18.68  &  406 &   82.0 &   14.0 \\
207.16310.1050 &  1.47  &    841 &   19.17 &   885  &  18.85  &  530 &   90.0 &   8.5 \\
207.16316.446 &  0.56  &    809 &   18.63 &   880  &  18.44  &  590 &   66.0  &  12.5 \\
208.15920.619 &  0.91  &    836 &   19.34  &  759  &  19.17  &  621 &   70.0  &  13.0 \\
208.16034.100 &  0.49  &    875 &   18.10 &   259  &  17.81  &  742 &   82.0  &  9.0 \\
211.16703.311 &   2.18  &    733 &   18.91 &   760  &  18.56  &  374 &   90.0 &   6.5 \\
211.16765.212 &   2.13  &    791  &  18.16 &   232  &  17.87 &   447  &  76.0 &   13.0 \\
1.4418.1930 &  0.53  &    960   & 19.61  &  340   & 19.42  &  577   & 64.5 &   11.5 \\
1.4537.1642 &  0.61  &    1107  & 19.31  &  367  &  19.15 &   635  &  79.0  &  9.0 \\
5.4643.149 &  0.17  &    936  &  17.48  &  943  &  17.15 &   699  &  80.0  &  9.0 \\
6.7059.207 &  0.15  &    977  &  17.88  &  392  &  17.41  &  613  &  83.0 &   11.5 \\
13.5962.237 &  0.17   &   879  &  18.95  &  899  &  18.47  &  524  &  66.0 &   12.5 \\
14.8249.74 &  0.22  &    861  &  18.90  &  444  &  18.60  &  579  &  69.0   & 13.5 \\
28.11400.609 &  0.44  &    313 &   19.61 &   321 &   19.31 &  504 &   70.5  &  13.5 \\
53.3725.29 &  0.06  &    266  &  17.64  &  249  &  17.20  &  553  &  69.5    & 14.0 \\
68.10968.235 &  0.39  &    243 &   19.92  &  261 &   19.40 &   566  &   83.0  &  12.0 \\
69.12549.21 &  0.14  &    253  &  16.92  &  244  &  16.50  &  497   &   68.0 &   12.5 \\
70.11469.82 &  0.08  &    243  &  18.25 &   241  &  17.59  &  544   & 79.0  &  10.5 \\
82.8403.551 &  0.15  &    836  &  18.89 &   857  &  18.55 &   556  &  56.5 &   12.0 \\
\hline
\end{tabular}
\end{table*}

A further complication is that for a few quasars, more than 4 peaks were
found. Our procedure to identify peaks begins with an autocorrelation
estimate on the semi-periodicity corrected data, followed by a smoothing
with a 50-day running-boxcar smoothing algorithm. All peaks having
autocorrelation greater than 1\% were tabulated for
processing in the orientation
angle measuring routine. For one quasar, 6 peaks were found in the V data
and 5 in the R data. So one peak was eliminated from the sextuplet and
another peak that was out of sequence was removed from both filter
solutions. For an additional 6 quasars, a fifth peak was found, but
comparison of the V and R filter data identified the spurious peak.

A final complication relates to the many quasars for which we determined an
orientation angle of approximately 60 degrees. For 6 objects we found the
pattern of equally spaced pulses corresponding to 60 degrees orientation,
but this could result from an artifact in the autocorrelation
function. Recall that if a secondary pulse occurs in a data stream, the
autocorellation produces a response at $t_1$ and $t_2$ but also at $t_2 - t_1$,
though at lower amplitude because the initiating central pulse is always a
factor 2 or 3 larger than the reverberations.
Thus we must presume that some of the autocorrelation peaks were biased
toward $t_2 - t_1$, $t_3 - t_1$, $t_3 - t_2$ etc. and possibly a false
peak created. As a further complication, these systematic problems arise
for just that part of the quasar reverberation diagram where the
inclination angle is a strong function of the first and second pulse
locations, meaning that the calculation is very sensitive to small
systematic errors.

For these regions, we have assumed that the angle determined for our
machine processing is an upper limit to the actual inclination angle, though
in our final results table we have entered the angles calculated. We plan
to take this uncertainty into account in our final
analysis of the measured inclination angle in the subsequent sections of
this report.

For MACHO quasar 206.17057.388 we find three autocorrelation peaks in
the R-filter data but only one in the less complete V data. We conclude
that the V-filter data are insufficient to give a solution and use only our
solution from R data.

Fig. 2 is a plot showing our estimated inclination angles, $\theta$ as a
function of the measured lags.  The solid curves are the predictions for Elvis
outflow surfaces with an internal structure variable $\epsilon=12^o$.  The
first reverberation peak is plotted with a plus symbol, the second with an
asterisk, third with a diamond and fourth with a triangle.  Notice the strong
agreement between the model's predictions and the observed reverberation
patterns.

\begin{figure*}
\begin{center}
\includegraphics{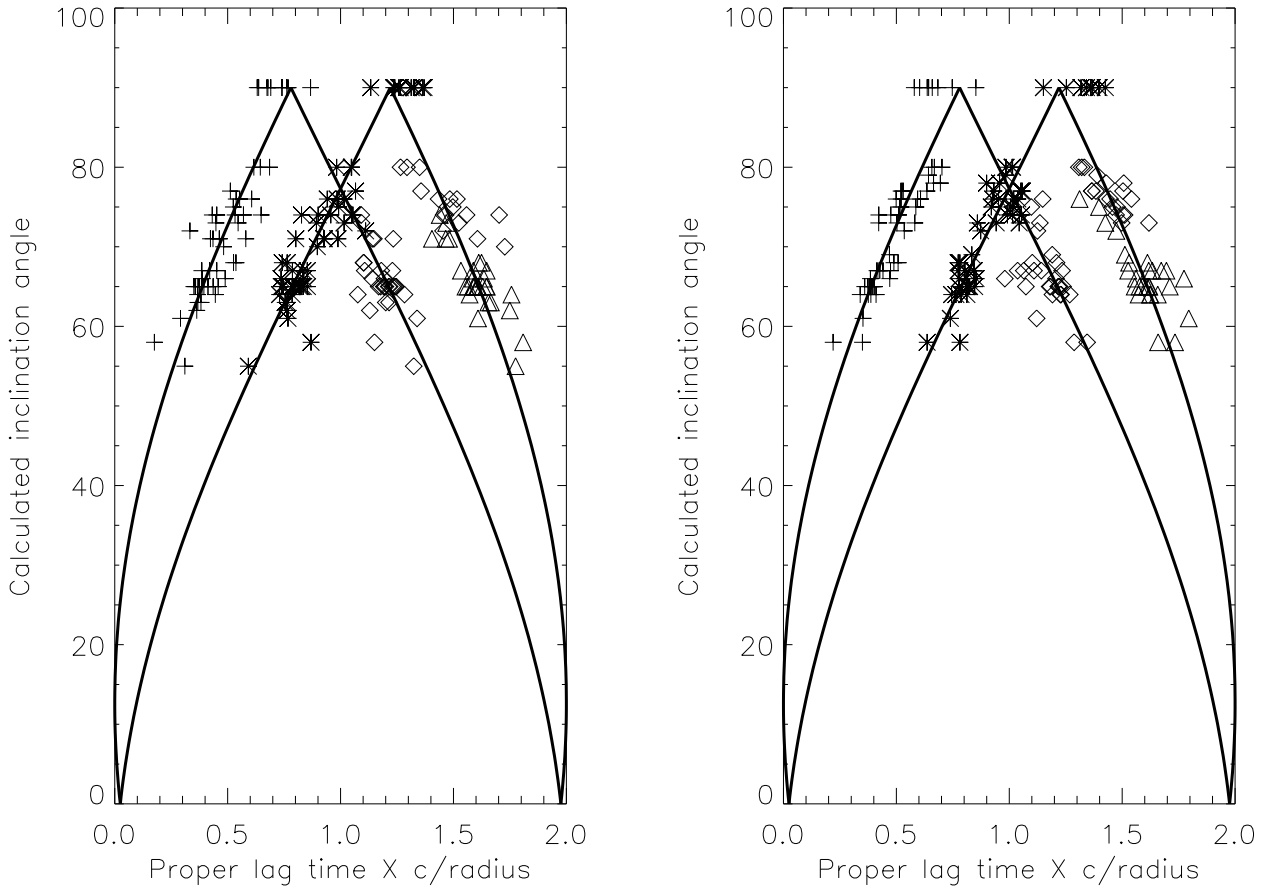}
\caption{A plot showing the order of reverberation pulses expected,
as a function of the viewing angle (between the plane of the sky and the
axis of rotation) of the quasar, for the V data (left) and for the R data
(right). The lines show the pattern of reverberations expected for the
reflected or fluorescing of radiation off of a centrally illuminated dusty
torus geometry. For inclinations near 0 and 78 degrees the pattern becomes
degenerate and the plotted points are limits, as described in Section 4.
For quasars viewed along their equators
the viewing angle is 0 and the initial pulse is broader because it occurs at
the same time as the reverberations from the nearest surfaces. }
\label{fig. 2}
\end{center}
\end{figure*}

\section{A Red Noise Simulation of our Inclination Angle Measurement
Procedure}

In Paper 1 we have already shown that for Gaussian noise, a noise
simulation can easily be distinguished from our actual quasar brightness
records. However our present procedure that determines the inclination
angles from autocorrelation peaks is essentially an extremely strong filter
that can mimic autocorrelation peaks with separations approximately the
same as found in our Fig. 2.

For this reason we have performed a simulation for a red noise component
that is related to Brownian motion noise. We created 10 sets of 55
simulated quasars
and with noise amplitude the same as our V and R filter quasar data. We
found that the red noise simulations were qualitatively different in that
they did not show consistently a pattern of reverberations like our MACHO
quasar data.

Thus in our real quasar data, we did not find any quasars that did not have
any significant reverberation structure, where significant is calculated on
the basis of the basic autocorrelation statistic that the noise related to
autocorrelation detection
is proportional to $\frac{1}{\sqrt{2N}}$ where N is the number of data
points in an individual brightness record.
However in our simulated data, 17 \% of brightness records showed
no autocorrelation peaks. This mean value 17 had an rms deviation among the
simulations of 6.3 so that an estimate of the mean and rms deviation is
$17 \pm \frac{6.3}{\sqrt{N-1}}$. In this way, our real data with 0 
autocorrelation
failures is an 8 sigma departure from the simulations. We conclude that our
procedures and results are not compatible with red noise (Brownian motion
noise) simulations.

\section{Quasar Properties Related to Inclination Angle}

With reasonable inclination angles available we
are in a position to look for other measured quasar properties that might
be related to this angle of observation, since such correlations might give
clues about the quasar's structure.

We first plot in Fig. 3  the measured lag to the autocorrelation minimum
(the anti-correlation peak).
The lag to this minimum is expected to be the full width of the primary
pulse at its base; this would be described as full width at profile base.

\begin{figure*}
\begin{center}
\includegraphics{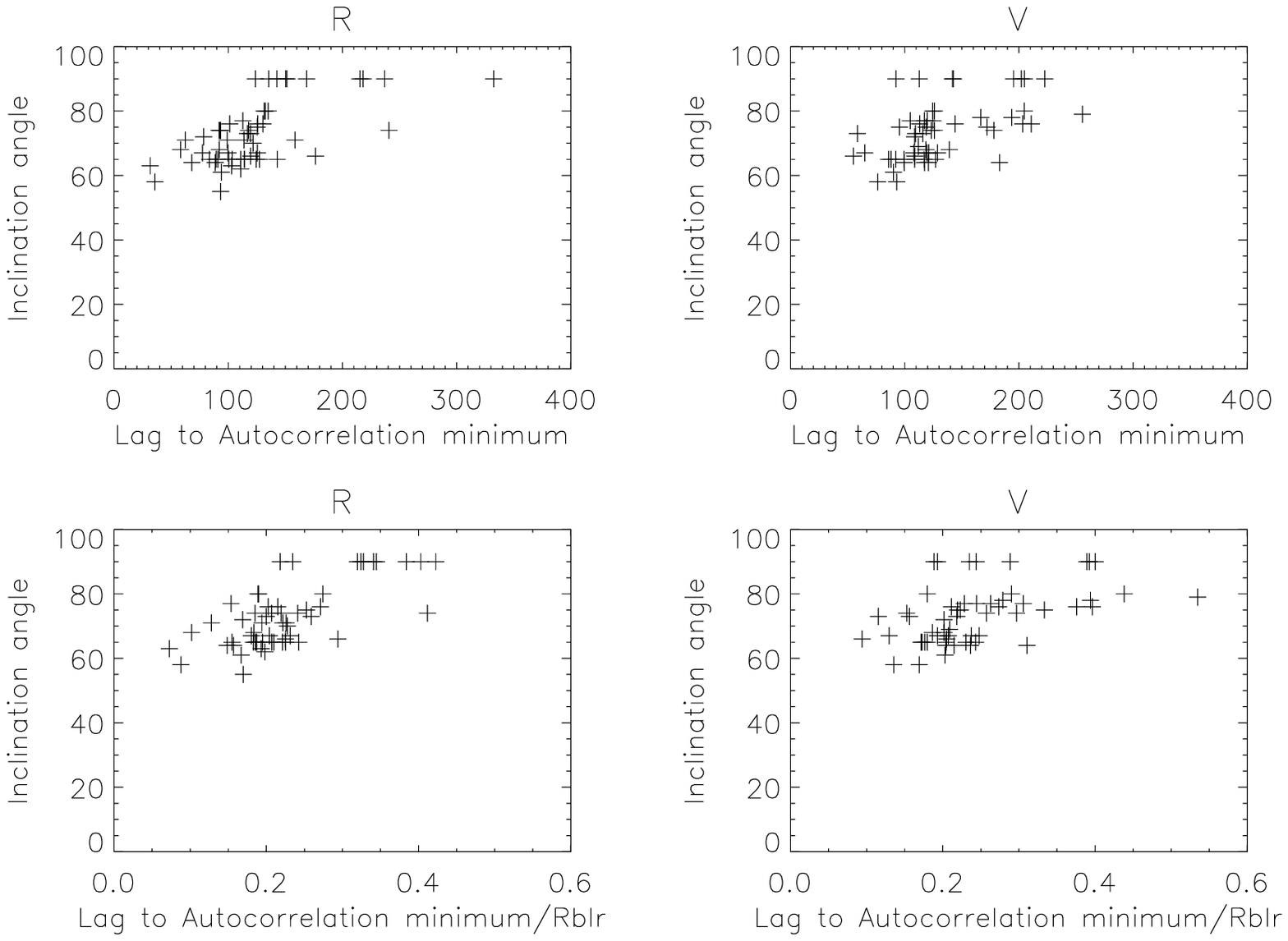}
\caption{Lag times to the anticorrelation peak for the R
  data (left) and the V data (right). The upper plots show the lags
  measured in days, whereas the bottom plots show the lags normalized to
  the radius of the broad-lined region, $r_{blr}$. }
\label{fig. 3}
\end{center}
\end{figure*}

We see in Fig. 3 that a significant correlation is evident. In the
top two panels, we present the correlation of the observed lags (in days) with
measured inclination, and in the bottom we show the lags normalized to the
$R_{blr}$ determined for the quasar for both filters. Thus for the two bottom
plots, the lags for the anti-correlation feature are divided by the lag of
$R_{blr}$. The appearance of correlation is seen in all four plots.

The existence of such a structure variable correlated with the
inclination angle is an important indicator of the nature of the central
structure. Since a spherically symmetrical inner quasar structure
should produce no angular dependence, given that the emitting region is
presumed to be outside of the region where strong general relativistic
beaming effects are expected, it may be immediately concluded from Fig. 3
that the inner luminous structure is not spherically symmetrical. This
conclusion is compatible with the disc-jet inner structure model normally
adopted.

We presently understand the anti-correlation peak to be associated with the
weak inter-pulse emission originating in a dark region of the accretion
disc. Such a dark region is expected to occur physically outside of the
luminous accretion disc inner edge which would be located near 6 $R_g$
and was measured to be near $4 R_g$ (Vakulik et al, 2007). This inner
luminosity was also measured to be $\frac{1}{3}$ of the Q2237 quasar's total.

For quasars seen nearly pole-on, the lag to the anti-correlation peak
represents the lag when the initiating central pulse has passed and the
reverberation pattern has not yet started. So the time lag to this minimum
luminosity should be inclination sensitive, as observed.

\section {Correlation of Anti-correlation Amplitude and $R_{blr}$}

In Fig. 1 we have already shown the lack of correlation between the
amplitude of the maximum anti-correlation and its lag for the V and B
data sets. Do either of these
quantities correlate with the measured inclination angle?

In Fig. 4 (top) we plot the amplitude of the anti-correlation peak as a
function of inclination angle for the R and V data. No correlation is
evident, and we have undertaken no further statistical tests.

Also in
Fig 4 (bottom) we show the measured $R_{blr}$ as a function of inclination
angle. Since the radius to the broad-line region should result from
reverberation and not from observing angle, no correlation is expected, and
none is found. However, an important conclusion may be drawn from this
plot. Effectively all of the measured $R_{blr}$ values lie within the range 400
- 800 days in R and in range 350 - 700 days in V. Ignoring the few extreme
values, we conclude that all of the radio quiet MACHO quasars have broad
line regions which are approximately the same size.
A glance at fig. 4 shows that the distribution of quasar sizes
is centrally concentrated and approximately symmetrical, with a mean value
of 545 days and a rms deviation of only 85 days. This implies that to an
excellent approximation, any magnitude-limit selected quasar in the
radio quiet state can be
presumed to have a radius to its broad line region of $545 \pm 85$ days.

\begin{figure*}
\begin{center}
\includegraphics{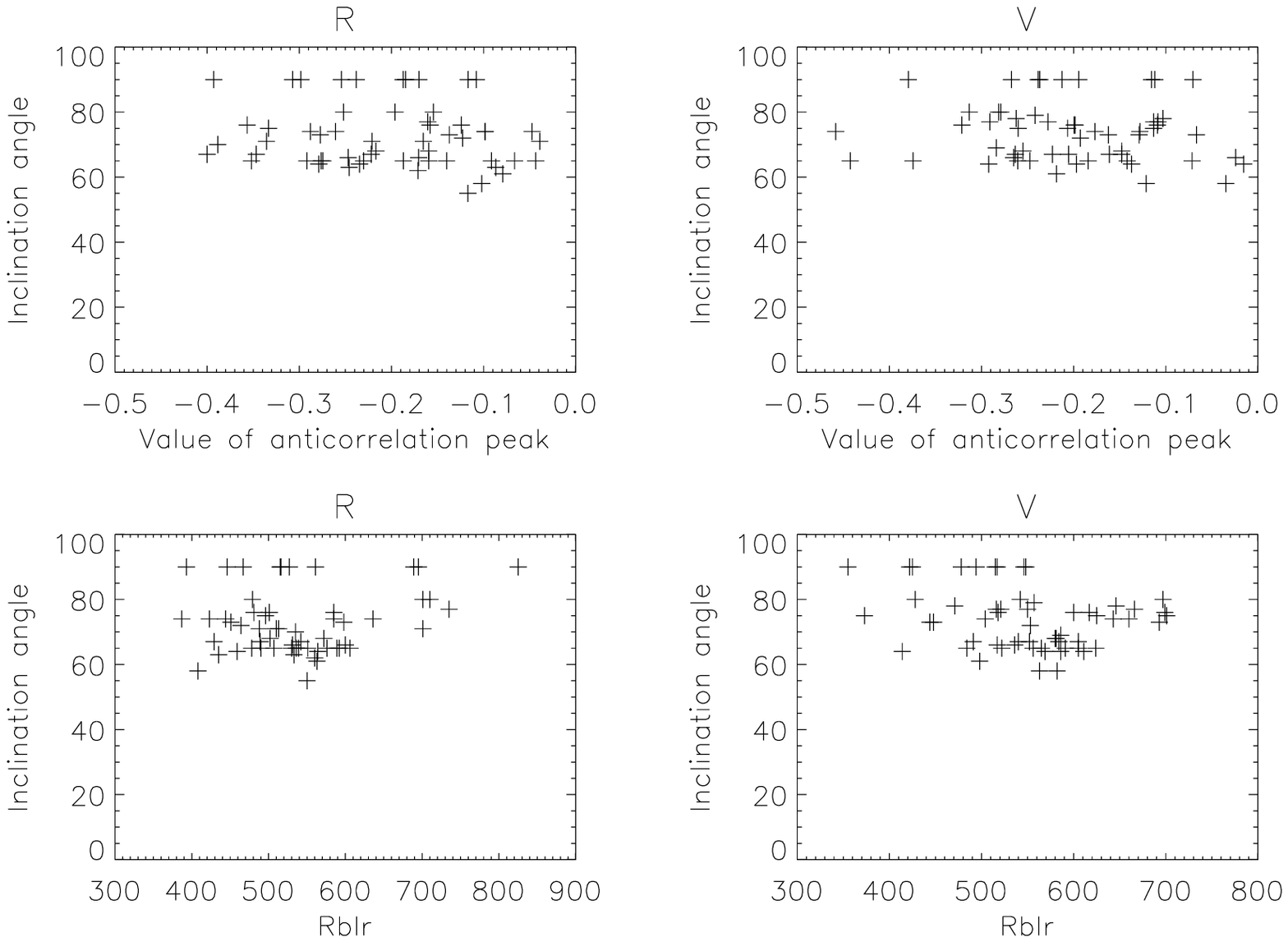}
\caption{ Correlation plots in the V filter (left) and R filter (right) for
  the value of the anti-correlation peaks (top) as a function of viewing angle.
  (bottom) The correlation of $R_{blr}$ as a function of viewing angle $\theta$.
  No correlation with viewing angle is found in any of the plots. However,
  notice the small variation of $R_{blr}$ values found, encompassing a total
  factor of two only, in the structure variable describing the overall
  quasar size. }
\label{fig. 4}
\end{center}
\end{figure*}

\section {The internal Structure Variable $\epsilon$}

Finally we discuss the internal quasar structure variable $\epsilon$, as
defined in Schild (2005). This is defined as the angle between the
accretion disc plane and the luminous portion of the Elvis outflow
surfaces, and is illustrated in Fig. 1 of Schild (2005). Previously we have
determined this angle to be 13 degrees for Q0957 (Schild, 2005) and also in
Q2237 (SLR08).

With $R_{blr}$ known for each quasar, it is simple
to compute $\epsilon$ from the 4 lag equations given in section 2 of
Paper 1. We show the $\epsilon$ values for the V and R filter data in Table 1.

The internal structure variable $\epsilon$ is best determined in the
quasars at intermediate inclination because of complex profile blending
effects at the extreme angles. Thus we determine the mean structure angle
for the 27 quasars with 4 reverberation peaks to be $12.0 \pm 0.5$ degrees
(1 -$ \sigma$ error of the mean).
We find that the mean value is satisfactorily close to the
value of 13 degrees previously determined for Q0957 (SLR06) and for
Q2237 (SLR08).

\section{Conclusions and Discussion}

In this report we have shown, from the analysis of 55 radio-quiet
MACHO quasars, that reverberation patterns in the brightness of
their UV-optical continuum indicate the presence of a universal
large-scale elliptical outflow structure, similar to that predicted
by the Elvis (2000) "dusty torus" model for quasars illuminated by
their central regions. Of course there might be other models that could
cause the pattern of brightness fluctuations found.

While it may yet be argued that the autocorrelation features observed may be
reproduced by some artificially tuned red-noise process, the fact that our
simulations predict a 17\% failure rate of red-noise to produce significant
autocorrelation structure leads us to conclude that noise is a highly unlikely
source of these patterns and that they are in fact real reverberation
processes. The combined conclusions from Paper 1 that reverberation
peaks originating at the region of the Elvis outflow
structures can be identified from the autocorrelation function, and also
can be seen by co-adding brightness record segments, suggests that they are
real and easily recognized and studied.  Similar structure
has been reported in the radio loud gravitationally lensed Q0957
quasar by SLR06. Since the broad blue-shifted high excitation
spectrum is a characteristic of all quasars, and since it is now
understood that the dusty torus with Elvis outflow wind creates the
emission lines, we expect the reverberation signature in brightness
records to be present in radio loud quasars, and therefore to be a
fundamental physical element of quasar structure that is easily studied
with reverberation.

The 55 radio quiet quasars in our sample showed a surprisingly small
dispersion in the various structural properties of the radius of
their broad line regions and of the polar opening angle of the
outflow wind with respect to their polar axis. In particular we
found that if a luminosity selected field quasar is observed to be
in the radio quiet state, an average broad-line region of 545 light
days size may be
adopted for it, accurate to within a 50 \% error. More importantly
the average of the polar opening angle of the outflow wind (the
complementary angle to the internal structure variable $\epsilon$ of
Schild (2005)) was found to be 78 degrees within a 3 degree
variation. The very large 78 degree polar opening angles observed
for the outflow winds in these quasars, which are similar to that
seen for the centrally driven magnetic outflows seen in cataclysmic
variables and young stellar objects, are difficult to explain in the
context of standard black hole accretion disk models for quasars.

Although many readers may find it surprising that our observations
of the 55 MACHO quasars show such small differences in inferred
reverberation region size, related to the broad-line region structure,
this result is not in contradiction currently known observations of
the difference between quasars and AGN. Quasars have long been known
to have very similar emission line spectra, and no elaborate system
of quasar spectral classification has emerged from 50 years of quasar
studies. This is to be contrasted to the case of AGN which includes
Seyfert galaxies that probably evidence galaxy interactions which
bring a great deal more complexity to study. For the Seyferts, an
elaborate classification system has in fact emerged. On the other
hand quasars have long been recognized to have large differences
in radio luminosity and high-energy X-ray emission, which have as
yet not been closely associated with significant difference in
their broad-line spectra.

\section{Acknowledgements}

We thank Pavlos Protopapas for help with accessing the MACHO program data
files, and we thank the MACHO Consortium for making the brightness records
publicly available. 
We thank Tesvi Mazeh and Victor Vakulik for
useful conversations about the properties of the autocorrelation
function. We also thank Phil Uttley for discussion of the properties
of red noise. Darryl Leiter thanks Alan Bridle of the National Radio
Astronomy Observatory (NRAO) for many insightful suggestions and for
his hospitality at NRAO during the 2008-2009 time period when this
research was performed.
This paper utilizes public domain data originally obtained by the MACHO
Project, whose work was performed under the joint auspices of the U.S.
Department of Energy, National Nuclear Security Administration by the
University of California, Lawrence Livermore National Laboratory under
contract No. W-7405-Eng-48, the National Science Foundation through the
Center for Particle Astrophysics of the University of California under
cooperative agreement AST-8809616, and the Mount Stromlo and Siding Spring
Observatory, part of the Australian National University.

\label{lastpage}

\end{document}